\documentclass[reprint,aip,amsmath,amssymb,jcp,superscriptaddress,longbibliography]{revtex4-2}
\usepackage{graphicx} 
\usepackage{dcolumn} 
\usepackage{bm} 
\usepackage[utf8]{inputenc}
\usepackage[T1]{fontenc}
\usepackage{etoolbox}
\usepackage{physics}
\usepackage{color}
\usepackage{hyperref} 
\usepackage{float}

\begin{document}
\title{A semiclassical non-adiabatic phase-space approach to molecular translations and rotations: A new picture of surface hopping and electronic inertial effects} 

\begin{abstract}
We present a novel semiclassical phase-space surface hopping approach that goes beyond the Born-Oppenheimer approximation and all existing surface hopping formalisms. 
We demonstrate that working with a correct phase-space electronic Hamiltonian can capture electronic inertial effects during pure nuclear translational and rotational motion and completely eliminate (at least to very high order) non-adiabatic transitions between electronic eigenstates. This work opens many new avenues for quantitatively investigating complex phenomena, including angular momentum transfer between chiral phonons and electrons as well as chiral-induced spin selectivity effects. 
\end{abstract}

\author{Xuezhi Bian}
\email{xzbian@princeton.edu} 
\affiliation{Department of Chemistry, Princeton University, Princeton, New Jersey 08544, USA}
\author{Yanze Wu} 
\affiliation{Department of Chemistry, Northwestern University, Evanston, Illinois 60208, USA}
\author{Tian Qiu} 
\affiliation{Department of Chemistry, Princeton University, Princeton, New Jersey 08544, USA}
\author{Tao Zhen} 
\affiliation{Department of Chemistry, Princeton University, Princeton, New Jersey 08544, USA}
\author{Joseph E. Subotnik}
\email{subotnik@princeton.edu}
\affiliation{Department of Chemistry, Princeton University, Princeton, New Jersey 08544, USA}
\affiliation{Department of Chemistry, University of Pennsylvania, Philadelphia, Pennsylvania 19104, USA}
\date{\today}

\maketitle 

\section{Introduction} \label{sec:introduction}
The Born-Oppenheimer (BO) approximation\cite{born1927} is perhaps the most fundamental concept in chemistry, stipulating that one can effectively freeze the nuclei when describe the electrons circulating in molecules and materials.
Mathematically, based on the fact that nuclei are much heavier than electrons, the molecular wavefunction is separated into a product of nuclear and electronic contributions.  As a practical matter, first, the electronic Hamiltonian is solved with clamped nuclei at different nuclear configurations to generate the ground  state electronic adiabats
 $\phi_j(\bm r; \bm R)$:
\begin{equation}
    \hat H_{\rm el} (\bm R) \ket{\phi_j(\bm r; \bm R)} = E_j (\bm R) \ket{\phi_j(\bm r; \bm R)}. 
\end{equation}
Second, the nuclear dynamics are propagated on the electronic potential energy surface. 
This approximation significantly simplifies molecular reactions and has led to numerous successes in predicting molecular properties using first-principle computational tools because, according to an exact Born-Huang expansion \cite{born1996}, the exact molecular wavefunction is a sum over many surfaces:
\begin{equation}
\Psi(\bm R, \bm r) = \sum_j \chi_j(\bm R) \phi_j(\bm r; \bm R), 
\end{equation}
where $\chi_j(\bm R)$ represents the nuclear wavefunction on the $j$-th adiabatic electronic surface.

\subsection{Curve Crossings vs. Electronic Inertial Effects}
Notwithstanding its many successes, the adiabatic BO separation described above can break down in various scenarios \cite{nelson2020}. 
When  nuclei move quickly enough, the inertia of electrons can prevent them from instantaneously following the nuclear motion, which of course in turn influences the nuclear dynamics as well. 
Now, although not very well appreciated, this breakdown of the BO approximation largely comes in two flavors. On the one hand, the BO approximation can fail dramatically in the close of vicinity of conical intersections\cite{yarkony1996,domcke2011} and curve-crossings\cite{landau1932,zener1932}, especially for long-distance electron transfer events when two electronic states come close together in energy\cite{gray2005}.  In analogy to electronic structure theory, one might call the above failures examples of ``static  correlation''\cite{helgaker2013}  between nuclei and electrons, and over the years, there has been a great deal of theoretical effort devoted to modeling such dynamics. On the dynamics front, there are now many different versions of Ehrenfest dynamics\cite{meyera1979,prezhdo1997} and surface hopping dynamics\cite{barbatti2011,subotnik2016}; on the electronic structure front, multi-reference electronic structure has been a major focus of advances over the last few decades\cite{lischka2018}.

On the other hand, and slightly less well-known, the BO approximation also loses accuracy in other settings even when two states are far apart, though the errors here are usually small or weak. In such a case, a perturbative analysis reveals that correcting a BO wavefunction requires small contributions for many different electronic states. 
In analogy with electronic structure theory, one might call these failures examples of ``dynamical correlation.''\cite{benavides2017}
On this front, there has been much less theoretical focus over the years, though there have been important advances -- most importantly in the small molecule community\cite{bunker1977,kutzelnigg2007, bubin2013}.
In the simplest case, one can understand these dynamical correlation effects of ``electronic inertia''--which is present even for the hydrogen atom. If one freezes the hydrogen nucleus (effectively ignoring the mass of the electron [as compared with the nucleus]), one finds that the electronic energy levels are of the form $E_n =  -\frac{m_e e^4}{8 \epsilon_0^2 h^2} \frac{1}{n^2}$. However, if one properly pulls out the center of mass of the hydrogen (and takes into account the finite mass of the electron), the result is slightly different, $E_n =  -\frac{\mu e^4}{8 \epsilon_0^2 h^2} \frac{1}{n^2}$.  Thus, in this case, accounting for electronic mass means simply replacing the raw electronic mass with the reduced mass $\mu = (M^{-1} +  m_e^{-1})^{-1}$ as far as the energy; more interestingly, however, this replacement also allows for electronic momentum, which is always lacking in standard BO theory\cite{nafie1983}.

Beyond the hydrogen atom, 
problems in dynamical nuclear-electronic correlation have been easily identified by high-resolution rovibrational spectroscopy for small molecules\cite{pavanello2012,wu2013}. As one might expect, 
early theoretical attempts to address this drawback introduced a mass correction to the nuclear mass that arises from perturbation theory, which works well for diatomic and triatomic molecular systems\cite{bunker19772,schwenke2001}.  However, these approaches are difficult to generalize for realistically sized molecules, especially in a fashion that matches a given electronic structure level of theory. 
Notably, \citeauthor{scherrer2017} derived a more general and rigorous position-dependent mass correction based on the exact factorization approach, which is applicable to polyatomic molecules; future work will need to determine if such a  method can be applied to real-time molecular dynamics (though likely on in the adiabatic regime if at all).



More recently, the role of electronic inertial effects has attracted attention among a community of solid state chemists and physicists investigating chiral phonons. Over the last few years, experiments have suggested that chiral phonon modes, which carry angular momentum\cite{zhang2015}, exhibit large phonon induced magnetic fields in terms of electron spin polarization\cite{shin2018,cheng2020,luo2023,davies2024}. However, classical electromagnetic theory, which states that rotating charges generate magnetic fields, usually fails to explain these experimental results by several orders of magnitude. 
To date, a few effective theories have been proposed to explained such extraordinary spin-phonon interaction phenomenon\cite{juraschek2022,geilhufe2022,geilhufe2023,chaudhary2024}. 
For example, \citeauthor{geilhufe2022} have pointed out that the large effective phonon magnetic moment may originate from electronic inertia effects, appearing as a Coriolis-type coupling between electrons and the rotating nuclei:
\begin{equation} \label{eq:coriolis1}
    H_{\rm s-ph} \propto {\bm \omega}_{\rm n} \cdot {\bm J}_{\rm e}, 
\end{equation}
where ${\bm \omega}_{\rm n}$ is the nuclear angular speed and  ${\bm J}_{\rm e}$ represents the electronic angular momentum. 
The form of spin-phonon coupling in Eq.~\ref{eq:coriolis1} is quite intuitive as electrons should feel can extra Coriolis force in a non-inertial nuclear rotating frame (which was realized long ago)\cite{wick1948,steiner1994}. Unfortunately, however, extending this Coriolis-type coupling to  a complete understanding of spin-chiral phonons has failed for two reasons. First,  a Coriolis-type coupling is usually restricted by theorists to pure rotational motion, whereas nuclei may also undergo translational and vibrational motions simultaneously. 
Second, for spin-related problems, multiple electronic states are often close in energy, making other non-adiabatic effects as significant as (or likely much more significant than) electronic inertia effects. 
Overall, there is a strong need to develop new computational tools that extend the Coriolis force to realistic molecules and allow for \textit{ab initio} dynamics simulations capable of quantitatively handling the electronic inertia effects over a broad range of systems, all the way from small molecules to clusters or layers of materials with spin-orbit coupling. 
Note that, in the context of BO dynamics, one has to include an extra nuclear Berry force to conserve the total angular momentum \cite{bian2023};
and in the context of Ehrenfest dynamics, one important correction already derived by Takatsuka\cite{amano2005} and Krishna\cite{krishna2007} is that a non-Abelian Berry Curvature must be added to the nuclear dynamics, which is essential for maintaining momentum conservation\cite{tao2024:e}.

\subsection{Electronic inertial effects from a standard surface hopping perspective}\label{sec:fssh}


Although surface hopping simulations have historically been applied to study static correlation problems (with strong non-adiabatic effects in localized regions) rather than dynamic correlation problems (with weak non-adiabatic effects in delocalized regions),  the theoretical chemistry community has already indirectly learned quite a lot about electronic inertial effects within the context of a Tully's surface hopping calculation in a basis of BO states.  
Within the surface hopping scheme\cite{tully1990}, one propagates classical nuclear trajectories on a single (active) electronic potential energy surface (here labeled by $j$):
\begin{equation}
    \dot {\bm R} = \frac {\bm P} M, 
\end{equation}
\begin{equation}
    \dot {\bm P} = -{\bm \nabla}_{\bm R} E_j, 
\end{equation}
and propagates quantum electronic amplitudes $\{c_j\}$ for each nuclear trajectory according to the electronic Schrodinger's equation:
\begin{equation}\label{eq:dc}
    \dot {c_j} = -\frac {i} \hbar E_jc_j - \sum_k \dot {\bm R} \cdot {\bm d_{jk}} c_k. 
\end{equation}
Here, we define the derivative coupling vector as ${\bm d}_{jk} = \bra{\phi_j} \bm \nabla_{\bm R} \ket{\phi_k}$. 
Nuclear trajectories  stochastically hop to another surface (say, $k$) with rate 
\begin{equation}
g_{j \to k} = \max\left[2\Delta t \Re\left(\bm R \cdot \bm d_{jk}\frac{\rho_{kj}}{\rho_{jj}}\right), 0\right]
\end{equation}
based on their electronic amplitudes.. Here $\Delta t$ is the discrete propagation time interval and $\rho_{jk}= c_j^* c_k$.

At this point, a straightforward question arises: Are electronic inertia effects incorporated in the surface hopping approach?   
At first glance, the answer appears to be no, because classical nuclei  move on BO adiabatic surfaces for standard surface hopping dynamics (and BO theory ignores electronic inertia as described above). Nevertheless, it is well known that, if one runs a trajectory where all atoms translate together, surface hopping will still predict a nonzero hopping rate. 
Physically, a hop is predicted because, even if all of the atoms are translating together, there is no guarantee that the electrons are translating; and after all, the electrons have some inertia.  Thus, surface hopping would seem to include electronic inertial effects to some extent. Mathematically, all of this is summed up by the fact that the derivative coupling is nonzero in the direction of a translation: 
\begin{equation} \label{eq:dtransneq0}
    \sum_{I} d_{jk}^{I\alpha} \neq 0,
\end{equation}
where $I$ represents the atomic index and 
if $\alpha = x, y, z$ labels a Cartesian direction.  
Note that a similar condition holds for rotations,  
\begin{equation}\label{eq:drotneq0}
    \sum_{I\beta\gamma} \epsilon_{\alpha\beta\gamma} R^{I\beta} d_{jk}^{I\gamma} \neq 0.
\end{equation}
so that pure rotations can also lead to non-adiabatic transitions (and again there must be some electronic inertial effects).

Now, ironically, almost all previous work  in semiclassical surface hopping calculations has sought to {\em remove} any trace of electronic inertial effects from a non-adiabatic simulation\cite{fatehi2012,parker2019}.  After all, for an isolated atom, it is easy to show that there is no hopping if one removes the center of mass coordinate, suggesting that any hop must be superfluous. 
Finally, at the time of a hop, it is well known that 
the nuclear momentum should be rescaled along the direction of the derivative coupling vector. 
However,  it was long ago noticed that, according to Eq.~\ref{eq:dtransneq0} and Eq.~\ref{eq:drotneq0},  when hopping, such momentum rescaling will not conserve the total nuclear linear and angular momentum for a single trajectory. 
For all of these reasons, most chemists have  imagined that inertial effects must be something of an annoyance when it comes to surface hopping and non-adiabatic dynamics. 
In fact, over the last thirty years, simple methods have been derived (including by one of the authors\cite{fatehi2011,shu2020,athavale2023}) to remove the translational and rotational parts of the derivative coupling vector by introducing electron translation factors (ETF) and electron rotation factors (ERF):
\begin{equation}
    \bm d_{jk} \to \bm d_{jk} - \bm d_{jk, \rm ETF} -  \bm d_{jk, \rm ERF}
\end{equation}
By projecting away the messier parts of the derivative coupling, one would seemingly restore nuclear momentum conservation at a hop, forbid  
hopping from pure molecular rotations and translations, and effectively eliminate any and all electronic inertia effects.

\subsection{Path Forward and Outline}
In retrospect, and with a  clear vision of electronic inertial effects now more apparent, all of the attempts above to correct surface hopping now appear quite misguided. 
These approaches disregard important electronic inertial effects. And at the end of the day, surface hopping still runs along BO state and thus incorrectly conserves the nuclear (but not total) linear and angular momentum, a failure which is especially important for time-reversal and spin symmetry breaking systems\cite{bian2023}. 
If one wishes to improve upon Tully's surface hopping algorithm, clearly a new paradigm is needed (and with an eye on how to include electronic inertial effects).

To that end, in this paper, we will explore a recently proposed semiclassical phase-space formalism that parameterizes the electronic Hamiltonian by nuclear momentum and position, $\hat H_{\rm PS}(\bm R, \bm P)$, going beyond the BO formalism (where the electronic Hamiltonian is parameterized by position alone,  $\hat H_{\rm BO}(\bm R)$).  
We will show that such an approach for simulating real-time dynamics can correctly account for electronic inertia effects. This phase-space formalism can furthermore be easily merged with semiclassical non-adiabatic dynamics algorithms, creating a phase-space surface hopping approach that is applicable in both strongly and weakly non-adiabatic limits (or, in the language of quantum chemistry, systems with static or dynamic electron-nuclear correlations). 
Moreover, this phase-space approach adds only a small computational costs to a Hartree-Fock (HF) or density functional theory (DFT) calculation using  a linear combination of atomic orbitals (LCAOs), which should allow for first-principles calculations for realistic-sized systems. 
As such, the present paper should serve as a strong endorsement of a phase-space approach to electronic structure theory and a reminder that electronic structure packages should not fear to go beyond the BO approximation.

The paper is organized as follows: In Sec.~\ref{sec:ps}, we review the phase-space approach and highlight how and  why one can and should build an electronic Hamiltonian parameterized by both $\bm R$ and $\bm P$.  In Sec.~\ref{sec:theory}, we then demonstrate that phase-space methods account for electronic inertial effects  insofar as the method prevents an unphysical non-adiabatic transition when a molecular system undergoes pure translations and rotations (unlike the case of standard BO surface hopping). 
In Sec.~\ref{sec:example}, we next provide several numerical examples in model molecular systems to support our theory. Finally, in Sec.~\ref{sec:discussion}, we discuss future applications and directions for the phase-space approach.

\section{\label{sec:ps} Review of the semiclassical phase-space surface hopping approach}

\subsection{History}

The idea of semiclassical phase-space surface hopping formalism originates with Berry's superadiabatic basis\cite{berry1987,berry1990}. For a time-dependent Hamiltonian $\hat{H}(t)$, the system wavefunction can be expressed in the (zeroth-order) adiabatic basis:
\begin{equation}
    \ket{\Psi(t)} = \sum_j c^0_j(t) \ket{\psi_j^0(t)}
\end{equation}
where
\begin{equation}
    \hat H(t) \ket{\psi_j^0(t)} = E_j^0(t) \ket{\psi_j^0(t)}.
\end{equation}
Then the time-dependent Schrodinger's equation in the zeroth-order adiabats can be written as:
\begin{equation}\label{eq:dc0}
    \dot {c}_j^0 = -\frac {i} \hbar E_j^0 c_j^0 - \sum_k  T_{jk} c_k^0. 
\end{equation}
where
\begin{equation}
    T_{jk} = \bra{\psi^0_j} \frac {d} {dt} \ket{\psi_k^0}.
\end{equation}
Berry proposed that one can diagonalize the zeroth-order time-dependent Hamiltonian 
\begin{equation}
     H^1_{jk}(t) = E^0_{k} \delta_{jk} - T_{jk} 
\end{equation} 
to generate  first-order adiabats:
\begin{equation}
    \hat H^1(t) \ket{\psi_j^1(t)} = E_j^1(t) \ket{\psi_j^1(t)}.
\end{equation}
This diagonalization can then be performed iteratively again and again to generate even higher-order adiabats. 
In principle, motion along a superadiabatic basis should reduce the number of non-adiabatic transitions.

In 2008,  Shenvi\cite{shenvi2009} took a big step forward by suggesting that one could run Tully's non-adiabatic surface hopping algorithm along superadiabats instead of the usual adiabatic (Born-Oppenheimer) states.  In the language of Berry, Shenvi suggested a phase-space surface hopping (PSSH) formalism along eigenstates of Berry's first-order adiabatic Hamiltonian
(the zeroth-order basis corresponding to  BO coordinate-space adiabats).
Mathematically, the first-order adiabatic Hamiltonian (or what we now call simply a phase-space Hamiltonian) can be written as:
\begin{equation} \label{eq:HPS}
H_{{\rm PS}, jk} (\bm R, \bm P) =  E_j  + \sum_l  \frac {(\bm P \delta_{jl} - i\hbar\bm{d}_{jl} )(\bm P \delta_{lk} - i\hbar\bm{d}_{lk} )} {2M} .
\end{equation}
and depends on both classical nuclear coordinates and momentum. 
Diagonalizing the phase-space Hamiltonian leads to phase-space adiabats:
\begin{equation}
 \hat H_{\rm PS}  \ket{\psi_j^{PS}(\bm R, \bm P)} = E^{\rm PS}_j(\bm R, \bm P) \ket{\psi_j^{\rm PS}(\bm R, \bm P)},
\end{equation}
Shenvi suggested running nuclear dynamics  based on Hamiltonian's equations of motion along these surfaces:
\begin{equation}\label{eq:drps}
  \dot {\bm R} =  \bm \nabla_{\bm P}E_{j}^{\rm PS},
\end{equation}
\begin{equation}\label{eq:dpps}
  \dot {\bm P} =  -\bm \nabla_{\bm R}E_{j}^{\rm PS},
\end{equation}
and propagating the electronic amplitudes (for surface hopping) as well along these same phase-space adiabats:
\begin{equation}\label{eq:dcps}
    \dot {c_j} =  \sum_k   - \frac {i} \hbar \bra{\psi_j^{\rm PS}} \left(\hat {H}_{\rm el}  - i\hbar \frac {d} {dt} \right)\ket{\psi_k^{\rm PS}}  c_k. 
\end{equation}

As explained by Tully originally\cite{tully1990}, for surface hopping dynamics, the algorithm that minimizes the number of hops is usually the most accurate\cite{wang2016}; hence the name, ``fewest switches surface hopping''. The same urge to minimize the number of hops applies equally well to the electronic basis: if one can find a basis where non-adiabatic transitions strictly vanish, then the proposed coupled nuclear-electronic dynamics can be simulated with only one surface without any unnatural hopping,  effectively reaching the exact factorization limit\cite{abedi2010}. For this reason, given the construction of the phase-space basis, Shenvi had strong hopes that the PSSH algorithm would revolutionize our framework for understanding non-adiabatic motion.  
However, in practice, Shenvi's PSSH suffers from three problems. First, propagating nuclear dynamics on phase-space adiabats is expensive, as it requires one to calculate derivatives of the derivative couplings when evaluating a  nuclear force at each time step.  Second, the algorithm is not numerically stable because the phase-space adiabatic energy diverges in the vicinity of conical intersections. Third and finally, the derivative coupling is not well-defined when the adiabatic surfaces are degenerate, so the algorithm cannot be applied to problems with spin and Kramers' degeneracy. 
Therefore, despite its conceptual strengths and empirical successes in several model Hamiltonians\cite{gherib2016}, as far as we are aware, the phase-space Hamiltonian in Eq.~\ref{eq:HPS} has never been implemented for first principle simulations.   

\subsection{A new phase-space electronic Hamiltonian}
Following Berry's seminal work on superadiabats and Shenvi's surface hopping approach, 
we have now demonstrated that one can build several classes of phase-space \textbf{electronic} Hamiltonians with different approximations for the derivative coupling in Eq.~\ref{eq:HPS}\cite{tao2024}, and all in such a way that 
the approximated derivative coupling can be included prior to diagonalization (so that there is only one diagonalization, as in standard BO theory).
These approaches have already led to promising results in the calculations of vibrational circular dichroism spectrum\cite{duston2024} as well as better vibrational energies in systems with ``heavy'' electrons and/or ``light'' nuclei\cite{bian2024}.

To define our new phase-space Hamiltonian, let us consider a mean-field LCAO-based electronic structure routine where the electronic Hamiltonian (or the Fock operator) can be expressed by a set of one-electron AO basis $\ket{\mu},\ket{\nu}$ with matrix elements: 
\begin{equation} 
H_{{\rm el},\mu\nu} = \bra{\mu} \hat H_{\rm el} \ket{\nu}. 
\end{equation}
In such framework, a simple guess for the phase-space electronic Hamiltonian is:
\begin{equation} \label{eq:HPSgeneral}
H_{\rm PS, \mu\nu} = H_{\rm el, {\mu\nu}} + \sum_{\sigma} \frac{(\bm P\delta_{\mu\sigma} - i\hbar{\bm \Gamma}_{\mu\sigma}) (\bm P\delta_{\sigma\nu} - i\hbar{\bm \Gamma}_{\sigma\nu})}{2M}.
\end{equation}
where ${\bm \Gamma}_{\mu\nu} = \bra{\mu}{\hat {\bm \Gamma}} \ket{\nu}$ is the key quantity representing the approximate ``derivative coupling''  introduced in the Hamiltonian. 
Then, by solving the phase-space electronic Hamiltonian in Eq.~\ref{eq:HPSgeneral}, one can obtain the phase-space adiabatic molecular orbitals (MO):
\begin{equation}
    \sum_{\nu} H_{\rm PS, \mu\nu} \phi_{j\nu} = \sum_{\nu} E_j^{\rm PS} S_{\mu\nu} \phi_{j\nu}
\end{equation}
where $S_{\mu\nu} = \bra{\mu} \ket{\nu}$ represents the overlap matrix between AO basis and $\phi_{j\nu}$ represents the MO coefficient. 

How should we choose the $\hat {\bm \Gamma}$ operator?  

\subsubsection{Translations}
Clearly, one intuitive choice for $\hat {\bm \Gamma}$ is to use the derivative between AO basis vectors:
\begin{equation} \label{eq:gammamunu1}
    {\bm \Gamma}_{\mu\nu} \overset{?}{=}  \bm d_{\mu\nu} = \bra{\mu}{\bm \nabla_{\bm R}} \ket{\nu}.
\end{equation}
After all, if the evolution of AO basis functions is accounted for, one effectively solves the electronic Hamiltonian in a nuclear moving frame \cite{artacho2017} where electronic inertia effects should not exist. 
Unfortunately, however, there are two problems making Eq.~\ref{eq:gammamunu1} not practical, and these need to be addressed.

The first problem is that ${\bm d}_{\mu\nu}$ is not anti-symmetric due to the non-orthonormality of AO basis. The phase-space Hamiltonian  is therefore not strictly Hermitian and becomes problematic to solve numerically. 
To fix this problem, we can choose to anti-symmetrize the AO derivative coupling by considering:
\begin{equation} \label{eq:gamma'}
     {\bm \Gamma}_{\mu\nu}' = \frac { {\bm d}_{\mu\nu} - {\bm d}^*_{\nu\mu}} 2 . 
\end{equation}
Note that ${\bm \Gamma}_{\mu\nu}' =  {\bm d}_{\mu\nu}$ when $\mu$ and $\nu$ are orbitals localized on the same atom.
Furthermore, one can show that the anti-symmetric component ${\bm \Gamma}_{\mu\nu}'$ carries the relevant information of  a basis evolving by molecular translation (see Appendix \ref{sec:gauge}). 
More importantly, it is easy to show that ${\bm \Gamma}_{\mu\nu}' $ satisfies:
\begin{equation} \label{eq:constrain1'}
 \sum_{I} -i\hbar  {\Gamma}_{\mu\nu}^{'I\alpha} + \bra{\mu} \hat p^\alpha_{\rm e} \ket{\nu} = 0,  
\end{equation}
which represents an important property for a derivative coupling-like term. 
The condition in Eq.~\ref{eq:constrain1'} is a
direct consequence of enforcing the phase convention\cite{littlejohn2023}: 
\begin{equation} \label{eq:convention1}
   \left( \hat {\bm P}_{\rm n} + \hat {\bm p}_{\rm e} \right) \ket{\mu} = 0, 
\end{equation}
which states mathematically  that the AO basis moves together with nuclear position.

\subsubsection{Rotations}
A second problem arises during molecular rotations. 
In principle, one would like to have a similar phase convention for rotations as in Eq.~\ref{eq:convention1} \cite{littlejohn2023}
\begin{equation} \label{eq:convention2}
   \left( \hat {\bm J}_{\rm n} + \hat {\bm J}_{\rm e} \right) \ket{\mu} = 0. 
\end{equation}
However, the AO basis in a typical electronic structure calculation is defined in the lab frame such that the basis functions do not rotate with the nuclei  when the molecule is rotating. 
Here, in order to capture also the effects of evolving basis during rotations, our solution is to construct an extra term: 
\begin{equation} \label{eq:finalgamma}
{\bm \Gamma}_{\mu\nu}   = {\bm \Gamma}_{\mu\nu}' + {\bm \Gamma}_{\mu\nu}'',
\end{equation}
by enforcing the convention in Eq.~\ref{eq:convention2}, i.e. 
\begin{equation} \label{eq:constrain2}
 \sum_{I\beta\gamma} -i\hbar \epsilon_{\alpha\beta\gamma} R^{I\beta} {\Gamma}_{\mu\nu}^{I\gamma} + \bra{\mu} \hat J^{\alpha}_{\rm e} \ket{\nu} = 0. 
\end{equation}
Note that the translation condition in Eq.~\ref{eq:constrain1'} must still be satisfied when including the $\bm \Gamma''$ term
\begin{equation} \label{eq:constrain1}
 \sum_{I} -i\hbar  {\Gamma}_{\mu\nu}^{I\alpha} + \bra{\mu} \hat p^\alpha_{\rm e} \ket{\nu} = 0,  
\end{equation}
 
In Appendix \ref{sec:ETFAO}, we give an explicit formula for a ${\bm \Gamma}_{\mu\nu}''$ matrix so that the total ${\bm \Gamma}_{\mu\nu}$ matrix satisfies
Eq.~\ref{eq:constrain2} and  
Eq.~\ref{eq:constrain1}. 

\subsubsection{A basis-free formulation}
While the phase-space electronic Hamiltonian approach described above (in Eqs.~\ref{eq:HPSgeneral} and \ref{eq:finalgamma}) relies on an AO  basis implementation, it is crucial to emphasize that the entire approach can in fact be extended to a completely basis-free formulation that does not depend on the specific properties of the AO basis\cite{tao2024:bf}. 
In this basis-free formulation,  the $\hat{\bm \Gamma}$ operator is defined by partitioning all of three-dimensional space according to the molecular geometry and masses of each nuclei. For concreteness, such a formulation is reviewed in the Appendix \ref{sec:bf}. 

Most importantly, in a basis-free implementation, the translational and rotational properties  of the matrix elements in Eqs.~\ref{eq:constrain2} and \ref{eq:constrain1} are enforced 
for the entire $\hat {\bm \Gamma}$ operator in the same manner. In other words, in a basis-free implementation of a phase-space electronic Hamiltonian, the Hamiltonian takes the form  
 \begin{equation}
 \hat H_{\rm PS} =  H_{\rm el} +  \frac{(\bm P  - i\hbar 
 \hat{\bm \Gamma})^2}{2M}.
 \end{equation}
and the $\hat{\bf \Gamma}$ operator satisfies:
\begin{equation} \label{eq:constrain1no}
 \sum_{I} -i\hbar {\hat \Gamma}^{I\alpha} +   \hat p^\alpha_{\rm e}   = 0,  
\end{equation}
\begin{equation} \label{eq:constrain2no}
 \sum_{I\beta\gamma} -i\hbar \epsilon_{\alpha\beta\gamma} R^{I\beta}  {\hat \Gamma}^{I\gamma} +  \hat J^{\alpha}_{\rm e}  = 0. 
\end{equation}
As we will show below, Eqs.~\ref{eq:constrain1no} and ~\ref{eq:constrain2no} (or AO basis equivalence Eqs.~\ref{eq:constrain2} and ~~\ref{eq:constrain1}) are the only constraints that we need to hold in order to find cancellation of non-adiabatic transitions under rotations and translations.  
Although the basis-free formulation represents a more general   (and numerically more stable) approach to the phase-space electronic Hamiltonian, in this manuscript, we will use the AO implementation for our results (which are quite easy to apply for very small systems).

\subsection{Successes}
Before concluding this section, let us make several remarks regarding the phase-space Hamiltonian in Eq.~\ref{eq:HPSgeneral}:

\begin{itemize}
    \item First, the phase-space Hamiltonian in Eq.~\ref{eq:HPSgeneral} is computationally efficient. It differs from the zeroth-order electronic Hamiltonian only by terms involving one-electron integrals ${\bm \Gamma}_{\mu\nu}$. 
    
    \item Second, the operator $\hat H_{\rm PS}$  is always well -defined, i.e., constructing this operator requires knowledge of only the  molecular geometry and the AO basis (and, in the basis-free version, one does not require knowledge of the AO basis either). Moreover, the Hamiltonian behaves smoothly at or near electronic degeneracies; it is not a perturbative Hamiltonian that blows up near degeneracies. 
    \item 
    Third, the new phase-space Hamiltonian is physically meaningful and incorporates how localized AO basis evolve with  moving nuclei during atomic translations and rotations.  
    As one would inevitably desire, the Hamiltonian depends only the relative (not absolute) position and orientation of the molecule. Mathematically, this fact is summed up by the fact the operator $\hat{\bm \Gamma}$ satisfies 
    \begin{equation} \label{eq:constrain3}
    {\bm \Gamma}_{\mu\nu}(\bm R_0) = {\bm \Gamma}_{\mu\nu}(\bm R_0 + \Delta\bm R) 
    \end{equation}
    and 
    \begin{equation} \label{eq:constrain4}
    {\bm \Gamma}_{\mathcal{R}\mu\mathcal{R}\nu} (\mathcal{R} \bm R_0) = \mathcal{R}{\bm \Gamma}_{\mu\nu} (\bm R_0).  
    \end{equation}
    where $\Delta \bm R$ represents a constant translation on all nuclei  and $\mathcal{R}$ represents a rotation operator in real-space. In a basis-free formulation, translational invariance can be expressed as
    \begin{equation}\label{eq:transjoe}
    \hat {\bm \Gamma} (\bm R, \hat{\bm r}, \hat {\bm p}, \hat {\bm s}) = \hat {\bm \Gamma} (\bm R + \Delta \bm R, \hat{\bm r} + \Delta \bm R, \hat {\bm p}, \hat {\bm s})
    \end{equation}
    and rotational invariance can be expressed as
    \begin{equation}\label{eq:rotjoe}
    \hat {\bm \Gamma} (\mathcal{R} \bm R , \mathcal{R}\hat{\bm r}, \mathcal{R}\hat {\bm p}, \mathcal{R} \hat {\bm s}) = \mathcal{R}\hat {\bm \Gamma} (\bm R, \hat{\bm r}, \hat {\bm p}, \hat {\bm s}) 
    \end{equation}
    
    \item Fourth, using Eqs.~\ref{eq:constrain1no}, \ref{eq:constrain2no}, \ref{eq:constrain3} and \ref{eq:constrain4},
    it is not difficult to prove that dynamics along a phase-space adiabatic surface conserve the total (nuclear + electronic) linear and angular momentum.\cite{tao2024} 
    
    \item Fifth, in the case of a single hydrogen atom, the phase-space Hamiltonian exactly reproduces the results from solving the electronic Hamiltonian in the molecular (electron + nucleus) center of mass frame, further suggesting that it provides more accurate solutions than the BO approximation.\cite{bian2024}

\end{itemize}

\section{Molecular Translation and Rotations in the Context of phase-space Surface Hopping \label{sec:theory}}

Let us now discuss non-adiabatic transitions within the  phase-space formalism, which is the central goal of this article. 
For this discussion, there is no need to worry about whether we work with single electronic states or many body states; all we need are the general equations in Eqs.~\ref{eq:constrain1no},~\ref{eq:constrain2no} to make the point.
Let us imagine that we diagonalize a phase-space Hamiltonian to generate a set of states  $\left\{\ket{\psi_j}\right\}$, so-called phase-space adiabatic states:
\begin{equation}\label{eq:psips}
    \hat H_{\rm PS} \ket{\psi_j} = E_j \ket{\psi_j}.
\end{equation}


Within a surface hopping framework, non-adiabatic transitions are described by the changes in quantum electronic amplitudes along  a classical nuclear trajectory. The time evolution of the quantum amplitudes follows Eq.~\ref{eq:dc}, such that: 
\begin{align}
    \frac {\partial c_j} {\partial t} &=\sum_{k}  -\frac i \hbar\bra{\psi_j} \left( \hat H_{\rm el} - i\hbar \frac d {dt}\right) \ket{\psi_k} c_k  \label{eq:dcdtps} \\
     &=  - \frac i \hbar E_j c_j - \sum_{k}  \frac i \hbar 
    \bra{\psi_j} \left( \hat H_{\rm el} - \hat H_{\rm PS}  \right) \ket{\psi_k}  c_k   \label{eq:dcdtps1}\\  
      & \quad - \sum_k \left( \dot{\bm R} \cdot \bra{\psi_j} \bm \nabla_{\bm R} \ket{\psi_k} + \dot{\bm P} \cdot \bra{\psi_j} 
     \bm \nabla_{\bm P} \ket{\psi_k}\right) c_k. \nonumber 
\end{align}   
The second term in Eq.~\ref{eq:dcdtps1} can be further expressed as:
\begin{equation}\label{eq:dcdt2}
\begin{aligned} 
&\quad -\sum_{k} \frac i \hbar 
    \bra{\psi_j} \left( \hat H_{\rm el} - \hat H_{\rm PS}  \right) \ket{\psi_k}  c_k \\
&= \sum_{k,l}  \left(\frac{\bm P} M   \cdot \bm \Gamma_{jk}  - i\hbar \frac {\bm \Gamma_{jl}\cdot \bm \Gamma_{lk}} {2M} \right)c_k.
\end{aligned}
\end{equation}
where $\bm \Gamma_{jk} = \bra{\psi_j} \hat {\bm \Gamma} \ket{\psi_k}$.

\subsection{Translations}
Consider now a molecular translation where the nuclear velocity $\dot{R}^{I\alpha}$ is a constant for all nuclei $I$ along direction the $\alpha$,   
and where the acceleration should remain zero, i.e, $\dot{P}^{I\alpha} = 0$. 
According to Eq. \ref{eq:transjoe}, the phase-space Hamiltonian is translationally invariant. From the translational invariance of the subsequent phase-space adiabats $\ket{\psi_j}$, i.e., 
\begin{equation}
   \bra{\psi_j} \left( \hat{\bm P}_{\rm n} +  \hat{\bm p}_{\rm e} \right) \ket{\psi_k} = 0,
\end{equation} 
the third term on the right hand side of Eq.~\ref{eq:dcdtps1} can be rewritten as:
\begin{equation}\label{eq:trans1}
\begin{aligned}
 &\quad - \sum_{kI\alpha} \left( \dot{R}^{I\alpha} \bra{\psi_j} \bm \nabla_{R^{I\alpha}} \ket{\psi_k} + \dot{P}^{I\alpha} \bra{\psi_j} 
 \bm \nabla_{P^{I\alpha}} \ket{\psi_k}\right) c_k\\
 &= \frac i \hbar \sum_{kI\alpha}  \dot{R}^{I\alpha} \bra{\psi_j} \hat p_{\rm e}^\alpha \ket{\psi_k} c_k.
\end{aligned}
\end{equation}
At this point, we can plug Eq.~\ref{eq:constrain1no} into Eq.~\ref{eq:dcdt2}, so that the first term on the right of Eq.~\ref{eq:dcdt2} becomes
\begin{equation}\label{eq:trans2}
\sum_{I\alpha}  \frac {P^{I\alpha}} M  \Gamma_{jk}^{I\alpha} =  -\frac {i} \hbar \sum_{I\alpha}   {\dot R}^{I\alpha}  \bra{\psi_j} \hat{\bm p}_{\rm e} \ket{\psi_k}\\ 
\end{equation}
Clearly,  Eq.~\ref{eq:trans1} cancels Eq.~\ref{eq:trans2}.
Thus, according to Eq. \ref{eq:dcdtps1}, $\dot {c_j} = 0 + \mathcal{O}(\hbar)$; in other words, there is no change of state to the zeroth-order in $\hbar$. Note that we have assumed $P^{I\alpha}/M = \dot{R}^{I\alpha}$ in Eq.~\ref{eq:trans2}. Clearly, this equality which is not strictly true in the phase-space formalism, but these two quantities  differ only by first and higher order $\hbar$ terms, so that the conclusion above still holds.

\subsection{Rotations}
Let us next turn to the case of molecular rotations.
For a molecule undergoing a rigid rotational motion with a constant angular velocity $\bm \omega$, one can express the nuclear momentum of the $I$-th atom as:
\begin{equation} \label{eq:angspeed}
   \frac  {{\bm P}^I} M  =  {\bm \omega} \times {\bm R}^I. 
\end{equation}
Within an AO basis formulation, the first-order $\bm \Gamma$ term in the phase-space Hamiltonian Eq.~\ref{eq:HPSgeneral} can be easily put into a familiar form:
\begin{equation}
\begin{aligned}\label{eq:coriolis2}
    -i\hbar \bra{\mu} \frac {\bm P} M \cdot \hat {\bm \Gamma} \ket{\nu} &= - i\hbar \sum_{I} \left( {\bm \omega} \times {\bm R}^I \right) \cdot   \bra{\mu} \hat {\bm \Gamma}^I \ket{\nu} \\
    &= -i\hbar \sum_I {\bm \omega} \cdot \left( \bm R^I \times  \bra{\mu} \hat{\bm \Gamma}^I \ket{\nu} \right) \\
    &=  -\bm \omega \cdot \bra{\mu} \hat {\bm J}_{\rm e} \ket{\nu}, 
\end{aligned}
\end{equation}
Here, we have used the rotational invariant condition for $\hat{\bm \Gamma}$ in Eq.~\ref{eq:constrain2}.  We find that 
the $\hat {\bm \Gamma} \cdot \bm P$ term recovers the Coriolis-type coupling in Eq.~\ref{eq:coriolis1} (in AO basis) which arises from the electronic inertia effects in rotating molecules, as suggested in Ref. \citenum{wick1948,steiner1994,geilhufe2022}.

Consider now the very general case of a rigid molecular rotation. Similar to Eq.~\ref{eq:angspeed}, for rigid molecular  rotations with angular speed $\bm \omega$, we express the change of nuclear momentum as:
\begin{equation}
    \dot{\bm P^I} = M (\bm \omega \times \dot{\bm R^{I}}) = \bm \omega \times {\bm P}^I.
\end{equation}
Then, we can rewrite the third term on the right hand side of  Eq.~\ref{eq:dcdtps1} as: 
\begin{equation}\label{eq:rot1}
\begin{aligned}
 &\quad \sum_{I} \left( \dot{\bm R}^{I} \cdot \bra{\psi_j} \bm \nabla_{\bm R^{I}} \ket{\psi_k} + \dot{\bm P}^{I} \cdot \bra{\psi_j} 
 \bm \nabla_{\bm P^{I}} \ket{\psi_k}\right) \\
 &= \sum_{I} ({\bm \omega} \times   {\bm R}^{I}) \cdot \bra{\psi_j} \bm \nabla_{\bm R^{I}} \ket{\psi_k} + ({\bm \omega} \times   {\bm P}^{I})  \cdot \bra{\psi_j} 
 \bm \nabla_{\bm P^{I}} \ket{\psi_k} \\
  &= \sum_{I} {\bm \omega} \cdot \left({\bm R}^{I}  \times \bra{\psi_j} \bm \nabla_{\bm R^{I}} \ket{\psi_k}) +   {\bm P}^{I}  \times \bra{\psi_j} 
 \bm \nabla_{\bm P^{I}} \ket{\psi_k} \right) \\
 &= -\frac i \hbar \sum_I \bm \omega \cdot \bra{\psi_j} \hat {\bm J}_{\rm e} \ket{\psi_k}. 
\end{aligned}
\end{equation}
Here, we have 
used the rotational invariance of the phase-space Hamiltonian (from Eq. \ref{eq:rotjoe}) which leads to rotationally invariant  phase-space adiabats satisfying:
\begin{equation}\label{eq:convention3}
\bra{\psi_j} \left( \sum_{I}   {\bm R^I} \times-i\hbar \bm\nabla_{\bm R^I} +  {\bm P^I} \times  -i\hbar \bm \nabla_{\bm P^I} + \hat{\bm J}_{\rm e}  \right) \ket{\psi_k} = 0. 
\end{equation}

Lastly, for the first term in Eq.~\ref{eq:dcdt2}, we have
\begin{equation}\label{eq:rot2}
\begin{aligned}
\quad \sum_{I}  \frac {\bm P^{I}} M  \cdot \bm \Gamma_{jk}^{I} 
&=  -\frac {i} \hbar \sum_{I}  \bm \omega \cdot \left(  {\bm R}^{I} \times     {\bm \Gamma}_{jk} ^I \right)\\
&= -\frac {i} \hbar \sum_{I} \bm \omega \cdot  \bra{\psi_j} \hat{\bm J}_{\rm e} \ket{\psi_k} .
\end{aligned}
\end{equation}
Again, if we  plug Eq.~\ref{eq:rot1} and Eq.~\ref{eq:rot2} into Eq.~\ref{eq:dcdtps1},  we find that non-adiabatic transitions vanish to first order  in $\hbar$ for a rigid molecular rotation.

The cancellation of non-adiabatic transitions under phase-space adiabatic basis during molecular translations and rotations reveals the power of a phase-space approach. 
In the context of surface hopping dynamics, fewer non-adiabatic transition almost always indicates better results.

\section{Numerical examples}\label{sec:example}

In this section, we present numerical examples to illustrate our theory above. 

\subsection{Traveling H-atom}
We begin with the simplest case of a traveling hydrogen atom. We initialized the hydrogen atom on its ground $1s$ state with a translational nuclear velocity of $\dot R = 1$ atomic unit (au) along the $z$-direction. 
For the electronic part, we truncated the electronic Hilbert space to include only the lowest five adiabatic electronic states (the exact $1s, 2s$ and $2p$ atomic orbitals). 
We then propagated the electronic Schrodinger equation     both in an adiabatic basis and a  phase-space adiabatic basis (defined by Eq.~\ref{eq:psips}), following Eq.~\ref{eq:dc} and Eq.~\ref{eq:dcdtps}, respectively. 
For the nuclear part, we employed the classical path approximation, assuming that the nuclear motion remains constant during propagation.

\begin{figure}[ht]
    \centering
    \includegraphics[width=\linewidth]{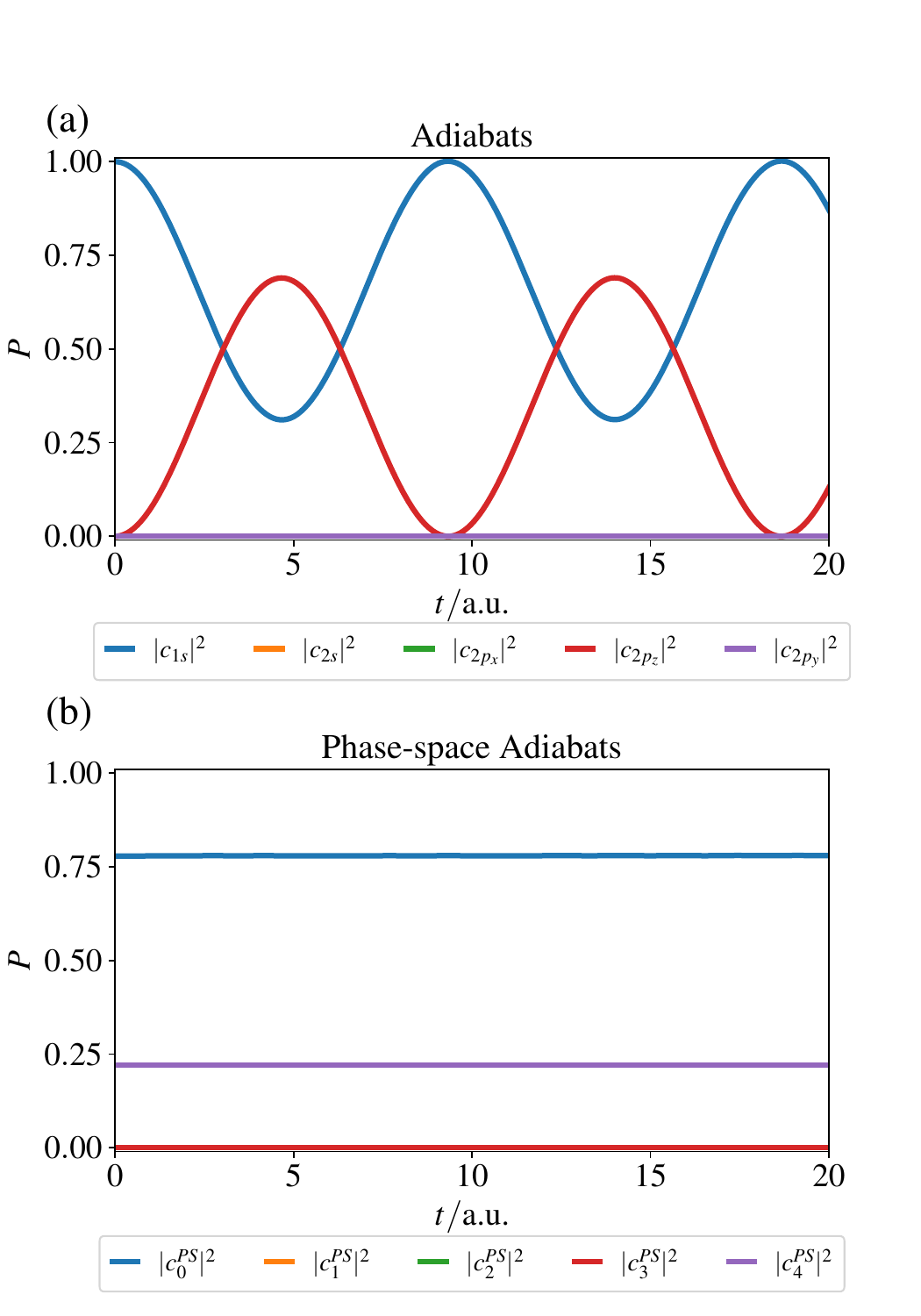}
    \caption{Electronic amplitudes for a hydrogen atom moving along the $z$-direction with a constant momentum of $\dot R = 1$ au in (a) the adiabatic basis and (b) the phase-space adiabatic basis.}
    \label{fig:Htrans}
\end{figure}

We plot the electronic amplitudes in the adiabatic basis in Fig.~\ref{fig:Htrans}(a). As expected, the electronic amplitudes oscillate dramatically between the $1s$ and $2p_z$ atomic adiabatic states due to the non-zero derivative coupling between them.
In Fig.~\ref{fig:Htrans}(b), in contrast, the electronic amplitude in the phase-space adiabatic basis remain constant during propagation.

 \subsection{Spin-Free Rigid Rotations of H$_2^+$}
Let us next consider the case of rigid rotations, focusing on a simple example: a $\rm H_2^+$ molecule. 
We employ the cc-pVDZ basis set for the $\rm H_2^+$ molecule and place the molecular ion in the $xy$-plane.  
We let the molecule  rotate around the $z$-axis with a constant angular velocity of $\omega = 1$ au. 
During the nuclear propagation, we keep the bond length a constant $R = 2$ au to eliminate rovibrational couplings.

In Fig.~\ref{fig:H2+rot}(a), we plot the electronic amplitudes in the adiabatic basis for the rigid rotating trajectory. 
The oscillating behavior of the adiabatic electronic amplitudes indicates continuous non-adiabatic transitions. In Fig.\ref{fig:H2+rot}(b) and (c), we present the electronic amplitudes in the phase-space adiabats, considering only $\hat {\bm \Gamma}'$ and both $\hat{\bm \Gamma}'$ and $\hat{\bm \Gamma}''$ in the phase-space Hamiltonian. 
We find that including only the translational $\hat{\bm \Gamma}'$ term does not eliminate the non-adiabatic transitions; it is necessary to also account for the rotation of atomic orbitals with respect to nuclear motion by considering the $\hat{\bm \Gamma}''$ term.

\subsection{Rotations and Spin}
\begin{figure}[H] 
    \includegraphics[width=\linewidth]{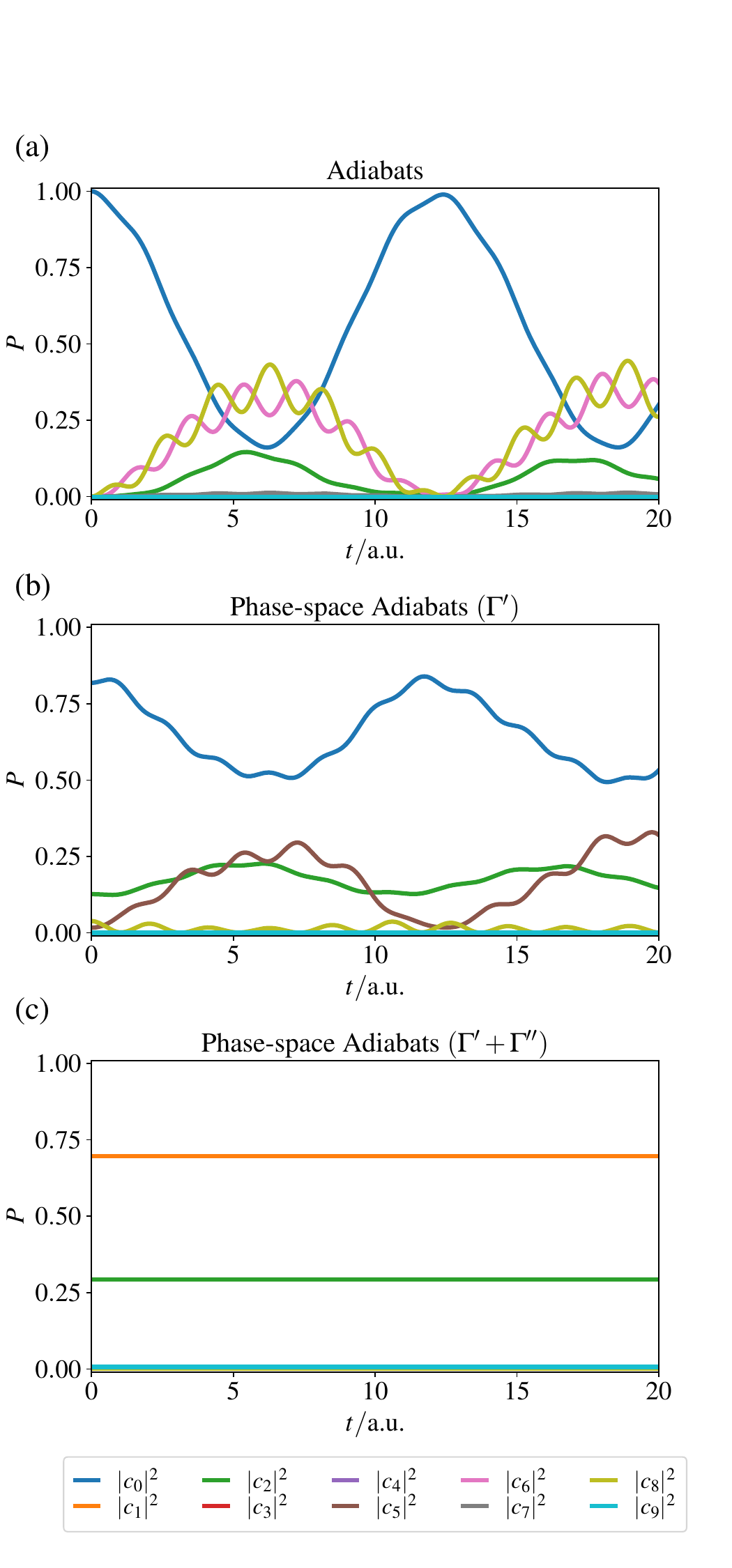}
    \caption{Electronic amplitudes of an $\rm H_2^+$ molecule during rigid rotational motion with a constant angular velocity of $\omega = 1$ au in (a) the adiabatic basis, (b) the phase-space adiabatic basis with only the $\hat{\bm \Gamma}'$ contribution, and (c) the phase-space adiabatic basis with both $\hat{\bm \Gamma}'$ and $\hat{\bm \Gamma}''$ contributions}
    \label{fig:H2+rot}
\end{figure}

\begin{figure}[H]
    \centering
    \includegraphics[width=\linewidth]{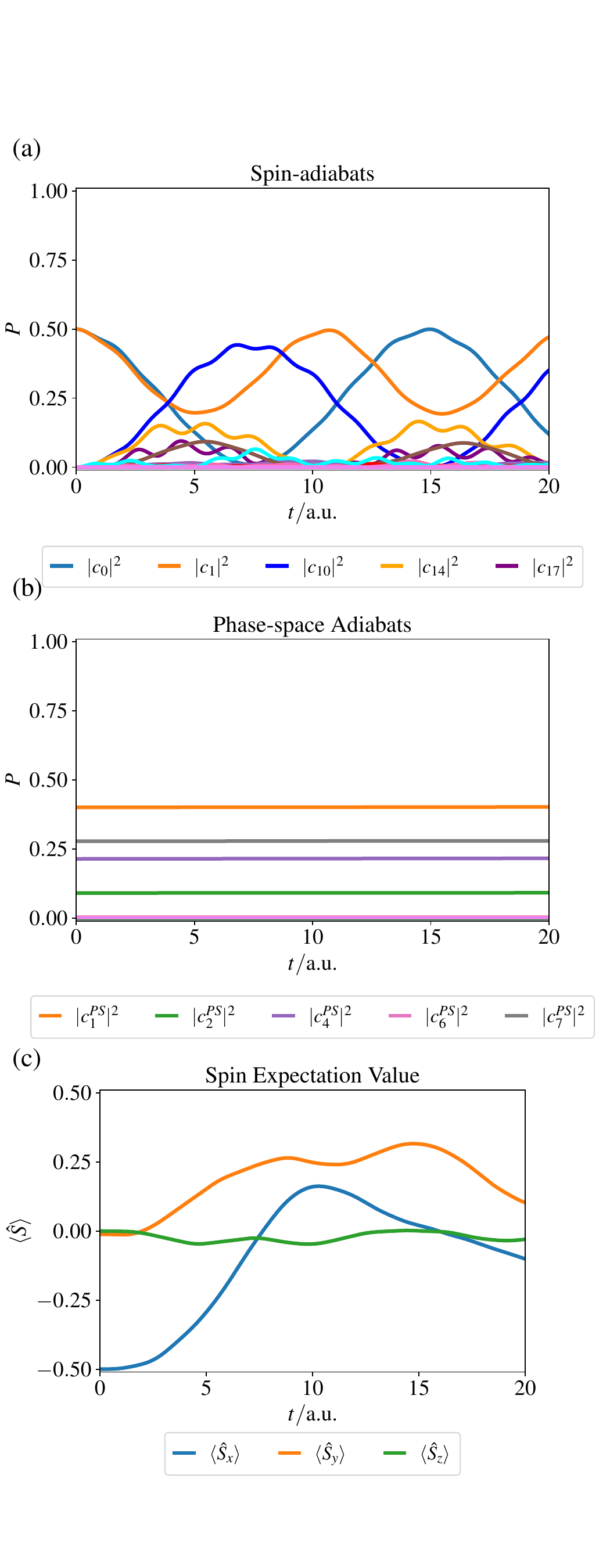}
    \caption{Electronic amplitudes of an $\rm H_2^+$ molecule with the same nuclear trajectory as in Fig.~\ref{fig:H2+rot} in (a) the spin-adiabatic basis and (b) the phase-space adiabatic basis, considering spin-orbit coupling. The expectation value of the spin is plotted in (c). In (a) and (b) only five most populated states are shown in the legends.}
    \label{fig:H2+rotsoc}
\end{figure}

Finally, let us turn to the case with electronic spin. 
To couple the spin DoFs with nuclei and electrons, we have included a one-electron Breit-Pauli spin-orbit coupling (SOC) operator in the electronic Hamiltonian:
\begin{equation}
    \hat H_{\rm SO} = \frac {\lambda \alpha_0^2} 2 \sum_{I} \frac {Z_I} {|\hat {\bm r} - {\bm R^I}|^3} \left((\hat {\bm r } - {\bm R}^{I}) \times \hat{\bm p})\right) \cdot \hat {\bm s}.
\end{equation}
with an additional  parameter $\lambda = 10^4$ to amplify the strength of the SOC.
We initialized the $\rm H_2^+$ molecule on the ground spin-adiabatic state with the spin pointing towards $x$ direction. 
Then we propagated electronic amplitudes with the same rigid rotational nuclear trajectory as in Fig.~\ref{fig:H2+rot}. 
In Fig.~\ref{fig:H2+rotsoc}(a), we show that the electronic amplitudes fluctuate in the spin-adiabatic basis as expected. In contrast, in Fig.~\ref{fig:H2+rotsoc}(b), the amplitudes in phase-space adiabatic basis remain constant. 
In Fig.~\ref{fig:H2+rotsoc}(c), the time-dependent spin expectation values indicate how the spin vector is ``rotating'' with nuclear motion. 

These results are very encouraging because non-adiabatic effects can be very important when considering spin DoFs.  
For example, for the $\rm H_2^+$ molecule discussed above (a system with an odd number of electrons), the spin-adiabats are at least two-fold degenerate according to Kramers' theorem.
In such case,  when the SOC is strong and the molecule is rotating (as shown in  Fig.~\ref{fig:H2+rotsoc}(a)), 
the expectation value of the spin must change physically, and within a surface hopping picture,  this change must be represented in a basis of spin-adiabats that is also rotating and changing.
Propagating such dynamics effectively is difficult for two reasons. On the one hand, to maintain angular momentum, one would like to invoke a Berry force but defining the on-diagonal Berry curvature within a pair of Kramers' degenerate surfaces requires an artificial gauge that can affect our results. On the other hand, it is unclear how to surface hop non-adiabatically between degenerate states in a unique way and we certainly want to avoid many hops back and forth; usually, fewer hops is better\cite{tully1990}.
Thus, a phase-space approach allows one to avoid many of the pitfalls of degeneracy that plague the surface hopping algorithm\cite{miao2019,bian2021,krotz2024}.


\begin{figure}[H]
    \centering
    \includegraphics[width=\linewidth]{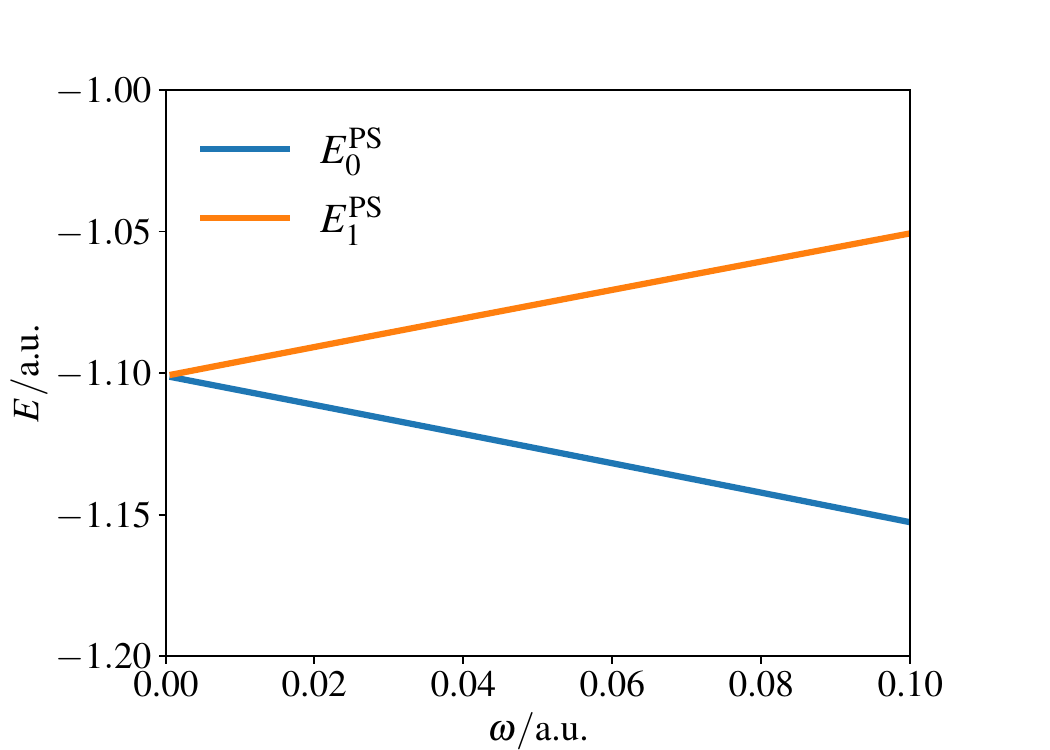}
    \caption{The lowest two phase-space energy surfaces as a function of angular speed $\omega$ for a $\rm H_2^+$ under rigid rotational motion.}
    \label{fig:gap}
\end{figure}

To emphasize just how strongly the phase-space formalism avoids the gauge problem, in Fig.~\ref{fig:gap}, we plot the lowest two phase-space adiabatic energies as functions of angular speed $\omega$ for the $\rm H_2^+$ molecule. 
Note that Kramers' degeneracy is lifted at non-zero nuclear velocities due to the Coriolis type coupling introduced in our phase-space formalism. 
Thus, single surface dynamics is straightforward in this formalism (and meaningful because one conserves angular momentum), and there is no problem applying a surface hopping algorithm.
 
\section{Discussion and Conclusions}\label{sec:discussion}
In this work, we have demonstrated, both analytically and numerically, the effectiveness of our new phase-space formalism in capturing electronic inertial effects across several molecular model systems. The semiclassical phase-space electronic Hamiltonian approach offers a new perspective on the fundamental translational and rotational motions of molecules, laying a solid foundation for its application to more complex real-time, non-adiabatic molecular dynamics. Furthermore, in a recent paper, we have shown that this phase-space approach also accurately captures electronic inertia effects in molecular vibrations, yielding significantly improved vibrational energy prediction\cite{bian2024}. Given these successes, we believe the phase-space approach has the potential to replace traditional semiclassical BO approaches in the near future.

Looking forward, there are many exciting directions that we aim to apply our phase-space approach to, where electronic inertia effects could play a significant role, especially in the condensed phase.
First, as mentioned in the introduction, one clear application is the spin-phonon problem, where our phase-space approach could be used to study the angular momentum transfer between nuclei and electrons through real-time non-adiabatic dynamics.
Second, the semiclassical phase-space approach can be extended to more general electron-phonon problems\cite{giustino2017}, as it offers a more accurate description of non-adiabatic effects. 
Such studies could indeed provide insight into the role of  non-adiabatic effects in phenomena like superconductivity\cite{frohlich1954}. 
Third and finally, by integrating the semiclassical phase-space formalism with surface hopping algorithms, we can explore electron transfer problems that involve spin DoFs. 
These simulation could offer a microscopic understanding of the chiral-induced spin selectivity (CISS) effect\cite{naaman2012,fransson2023,bloom2024} potentially resolving long-standing questions in this area.

 
\section{Acknowledgments}
This work was supported by the National Science Foundation under
Grant No. CHE-2422858 (J.E.S). X.B. thanks Titouan Duston for the helpful discussions.  

\appendix

\section{Molecular translations and the anti-Hermitian part of $d_{\mu\nu}$}
\label{sec:gauge}
Let us consider a molecule at position $\bm R$, where the overlap matrix elements between two AO basis functions are given by:
\begin{equation}
   S_{\mu\nu}(\bm R) = \bra{\mu (\bm R)} \ket{\nu(\bm R)}.
\end{equation} 
Now, suppose all nuclei in the molecule are translated by a displacement  $d\bm R$. 
We can expand the AO basis functions to first order with respect to this displacement as:
\begin{equation}
    \ket{\mu (\bm R + d\bm R)} = \ket{\mu (\bm R)} +  \ket{\frac {\partial \mu (\bm R)} {\partial \bm R}} \cdot \Delta \bm R
\end{equation}
Thus, the overlap matrix at the translated position $\bm R + \Delta\bm R$ becomes (to the first order): 
\begin{equation}
    S_{\mu\nu}(\bm R + \Delta\bm R) = S_{\mu\nu}(\bm R) + \left(\bm d_{\mu\nu} + \bm d_{\nu\mu}^* \right) \cdot \Delta\bm R.  
\end{equation}
Since the overlap matrix elements should remain unchanged under pure molecular translations, we have the condition:
\begin{equation}
    S_{\mu\nu}(\bm R + \Delta\bm R) = S_{\mu\nu}(\bm R). 
\end{equation}
which implies the following relation:
\begin{equation}
     \sum_I \bm d_{\mu\nu}^I + \bm d_{\nu\mu}^{I*} = 0.
\end{equation}
If we express the total derivative coupling between the AO basis in terms of its Hermitian and anti-Hermitian components, i.e.,  $(\bm d_{\mu\nu} + \bm d_{\nu\mu}^*)/2$ and  $(\bm d_{\mu\nu} - \bm d_{\nu\mu}^*)/2$, it becomes clear that only the anti-Hermitian component is relevant during molecular translations (Eq.~\ref{eq:gamma'}).

\section{Explicit definitions of the $\Gamma''$ term in AO basis}
\label{sec:ETFAO}
As stated in the main text, including only the anti-Hermitian components of the AO derivative coupling $\bm \Gamma'$ is insufficient for describing molecular rotations; an additional ERF  correction term $\bm \Gamma''$ must also be considered for conservation of angular momentum.
The explicit definition for $\bm \Gamma''$ that we employ  can be expressed in an AO basis as:
\begin{equation}
  \bm{\Gamma}''^{A}_{\mu\nu} = \zeta_{\mu\nu}^I\left(\bm{R}^I-\bm{R}_{\mu\nu}^0\right) \times \left(\bm{K}_{\mu\nu}^{-1}\bm{J}_{\mu\nu}\right) , 
\end{equation}
where we assume $\mu$ and $\nu$ are AO basis centered around $J$ and $K$, and 
\begin{equation}
    \zeta^I_{\mu\nu} = \exp\left(-\eta \frac{2|(\bm{R}^I-\bm{R}^J)|^2 |(\bm{R}^I-\bm{R}^K)|^2}{|(\bm{R}^I-\bm{R}^J)|^2 + |(\bm{R}^J-\bm{R}^K)|^2}\right) ,
\end{equation} 
\begin{equation}
    \bm{R}_{\mu\nu}^0 = \sum_I \zeta_{\mu\nu}^I\bm{R}^I /\sum_I \zeta_{\mu\nu}^I,
\end{equation}
\begin{equation}
\begin{aligned}
    \bm{K}_{\mu\nu}&=-\sum_{I}\zeta_{\mu\nu}^I\left((\bm{R}^I -\bm{R}_{\mu\nu}^0)^\top (\bm{R}^I-\bm{R}_{\mu\nu}^0)\right)\mathcal{I} \\ \quad &+\sum_I\zeta_{\mu\nu}^I(\bm{R}^I -\bm{R}_{\mu\nu}^0)(\bm{R}^I -\bm{R}_{\mu\nu}^0)^\top,
\end{aligned}
\end{equation}
\begin{equation}
   \bm{J}_{\mu\nu}= \frac{1}{i\hbar}\bra{\mu}{\frac{1}{2}\left(\hat{\bm{l}}^{(J)}+\hat{\bm{l}}^{(K)}\right) + \hat{\bm s}} \ket{\nu}. 
\end{equation}
Here $\zeta^I_{\mu\nu}$ represents a semi-local function that ensures the locality of $\Gamma''$, and $\eta$ is the parameter controls the degree of locality.  
In this manuscript, we did not apply any locality constraint, i.e., we set $\zeta^I_{\mu\nu} = 1$ as the molecular systems are relatively simple. 
In Eq.~\ref{eq:b5}, 
$\hat {\bm l}^{(J)} = (\hat {\bm r} - \bm R^J) \times \hat {\bm p}$ and $\hat {\bm l}^{(K)} = (\hat {\bm r} - \bm R^K) \times \hat {\bm p}$ are electron angular momentum operators around atoms $J$ and $K$. 
It is then straightforward to prove that $\bm \Gamma''$ satisfies the conditions in Eq.~\ref{eq:constrain2} and Eq.~\ref{eq:constrain4}. 
For further details and discussion, we refer readers to Refs.~\citenum{qiu2024,tao2024}. 

\section{Basis-free definition of the $\hat{\Gamma}$ operator}
\label{sec:bf}
More generally, one can also define a basis-free expression for the $\hat {\bm \Gamma}$ operator as follows:
\begin{equation}
    \hat {\bm \Gamma} = \hat {\bm \Gamma}' + \hat {\bm \Gamma}''.
\end{equation}
The translational component $\hat {\bm \Gamma}'$ is defined as:
\begin{equation}
    \hat {\bm \Gamma}'^I = \frac 1 {2i\hbar} \left(\Theta^I(\hat {\bm r}) \hat{\bm p} + \hat{\bm p}\Theta^I(\hat {\bm r})  \right)
\end{equation}
where $\Theta^I(\hat {\bm r})$ represents a space-partitioning operator that determines how much the electron is influenced by the momentum of nucleus $I$ based on the electron's spatial position. 
There are many ways to define the $\Theta^I(\hat {\bm r})$ operator, and one example is 
\begin{equation}
    \Theta^I(\hat {\bm r}) = \frac {M^I e^{-|\hat {\bm r} - {\bm R}^I|^2}/\sigma_I^2} {\sum_J M^J e^{-|\hat {\bm r} - {\bm R}^J|^2}/\sigma_J^2 }.
\end{equation}
Here $\sigma_I$ is a parameter that controls the locality of the momentum coupling.

Similarly, for the rotational component $\hat {\bm \Gamma}''$, we have 
\begin{equation}
  \hat{\bm{\Gamma}}''^{I}  = \sum_{JK} \zeta_{JK}^I\left(\bm{R}^I-\bm{R}_{JK}^0\right) \times \left(\bm{K}_{JK}^{-1}\hat {\bm{J}}^J\delta_{JK} \right) , 
\end{equation}
where 
\begin{equation}
    \zeta^I_{JK} = \exp\left(-\eta \frac{2|(\bm{R}^I-\bm{R}^J)|^2 |(\bm{R}^I-\bm{R}^K)|^2}{|(\bm{R}^I-\bm{R}^J)|^2 + |(\bm{R}^J-\bm{R}^K)|^2}\right) ,
\end{equation} 
\begin{equation}
    \bm{R}_{JK}^0 = \sum_I \zeta_{JK}^I\bm{R}^I /\sum_I \zeta_{JK}^I,
\end{equation}
\begin{equation}
\begin{aligned}
    \bm{K}_{JK}&=-\sum_{I}\zeta_{JK}^I\left((\bm{R}^I -\bm{R}_{JK}^0)^\top (\bm{R}^I-\bm{R}_{JK}^0)\right)\mathcal{I} \\ \quad &+\sum_I\zeta_{JK}^I(\bm{R}^I -\bm{R}_{JK}^0)(\bm{R}^I -\bm{R}_{JK}^0)^\top,
\end{aligned}
\end{equation}
\begin{equation}\label{eq:b5}
   \hat{\bm{J}}^I= \frac{1}{2i\hbar}\left(
   (\hat{\bm r} - \bm R^I) \times \Theta^I(\hat {\bm r}) \hat{\bm p} +  (\hat{\bm r} - \bm R^I) \times \hat{\bm p} \Theta^I(\hat {\bm r}) + 2\hat {\bm s} 
   \right)
\end{equation}
For more details, see Ref.~\citenum{tao2024:bf}.

\bibliography{main}
\end{document}